\documentclass[aps,twocolumn,showpacs,superscriptaddress,pra,10pt]{revtex4-1}
\usepackage{graphicx}
\usepackage{dcolumn}
\usepackage{bm}
\usepackage{color}
\usepackage{amsmath}
\usepackage{amsfonts}
\usepackage{subfigure}
\usepackage[makeroom]{cancel}

\newcommand{\COMMENTED}[1]{}

\begin{document}

\author{Ettore Vitali}
\affiliation{Department of Physics, California State University Fresno, Fresno, California 93740}

\author{Peter Rosenberg}
\affiliation{National High Magnetic Field Laboratory, Florida State University, Tallahassee, Florida 32310}

\author{Shiwei Zhang}
\affiliation{Center for Computational Quantum Physics, Flatiron Institute, 162 5th Avenue, New York, New York 10010}
\affiliation{Department of Physics, College of William and Mary,  Williamsburg, Virginia 23185}

\title{Calculating
ground state properties of correlated fermionic systems with BCS trial wave functions in 
Slater determinant path-integral approaches}

\begin{abstract}
We introduce an efficient and numerically stable technique to  make use of a BCS trial wave function
in the computation of correlation functions  of strongly correlated 
quantum fermion systems.
The technique is applicable to any projection approach involving paths of independent-fermion 
propagators, for example in mean-field or  
auxiliary-field quantum Monte Carlo (AFQMC) calculations. 
Within AFQMC, in the absence of the sign problem, the
methodology allows the use of a BCS reference state 
which can
greatly reduce the required imaginary time of projection, and improves Monte Carlo sampling efficiency and 
statistical accuracy for systems where pairing correlations are important.   
When the sign problem is present, 
the approach 
provides a powerful generalization of the constrained-path 
AFQMC 
technique which usually uses 
Slater determinant trial wave functions. As a demonstration of the capability of the methodology, we present benchmark results for the attractive Hubbard model, both spin-balanced 
(no sign problem) and with a finite spin polarization (with sign problem).
\end{abstract}

\maketitle

\section{Introduction}

Strongly correlated many-body systems are a central challenge of modern physics. These systems exhibit a variety of exotic phenomena in a wide array of physical contexts, from high-temperature superconductors to the cores of neutron stars.
Despite	the fundamental importance of strongly correlated systems, there are at present 
 relatively few quantitative theoretical treatments of these systems at the many-body level. One cutting-edge method that has demonstrated strong 
 capabilities in the study of ground states and excitations of strongly correlated many-body systems is auxiliary-field quantum Monte Carlo (AFQMC) \cite{CP-PhysRevB.55.7464,CP-FT-PhysRevLett.83.2777,phaseless2003}. This method formulates the
 calculation of ground-state properties as an imaginary-time propagation of 
 an entangled ensemble of
 independent-particle solutions in auxiliary-fields. The framework united key ingredients of
determinantal Monte Carlo \cite{BSS,AFQMC_Koonin,Assaad-lecture-notes-2002}
and real-space diffusion Monte Carlo (DMC) \cite{Ceperley-Alder_DMC,Ceperley_DMC_RMP},
casting the solution of the many-body ground state problem as an iterative process involving
mean-field states, as in standard density-functional theory (DFT) calculations, living in 
fluctuating external fields.
It is efficiently realized computationally as 
 open-ended, importance-sampled random walks in 
a manifold of Slater determinants. 

\COMMENTED{
The technique is routinely applied to quantum chemical systems \cite{Mario_AFQMC_QC} and to lattice models of interacting electrons \cite{PhysRevB.94.085103,Ettore_3Band,ettoreGAP}
and ultracold atoms \cite{Hao-2DFG,soc-qmc,AFQMC_Rashba_OPLATT}.  

Recent benchmark papers have shown that AFQMC is indeed a method of choice for a very large class of Hamiltonians in condensed matter physics and quantum chemistry \cite{hubbard_benchmark,h_benchmark}. It has also lead to great success in solving the problem of an attractive Fermi gas with zero-range interactions in two-dimensions \cite{Hao-2DFG,2DFG_Drut}.

The AFQMC technique relies on random walks within the manifold
of independent-particle solutions, 
which are wave functions describing independent fermions.
The method provides a stochastic linear combination of Slater determinants, the so-called {\it{walkers}},
 which are Monte Carlo samples of the ground state wave function of a given fermionic system. One could think of 
the usual stochastic dynamics underlying the traditional 
DMC 
as a special case, where the orbitals in the Slater determinants are 
made of eigenfunctions of the position operator \cite{CP-PhysRevB.55.7464}.
}

In certain situations the
AFQMC technique is sign-problem-free (as is the standard determinantal Monte Carlo approach), 
and numerically exact results can be obtained. 
There are many examples of important applications along these lines \cite{PhysRevB.94.085103,ettoreGAP,Sorella_NegU_Hubbard_2018,Sorella_TriangularHubbard_2018,Assaad_KMH_AFQMC_2012,Assad_KMCoulomb_2014,Yao_MQMC_2015,Yao_Majorana_AFQMC_2018,Scalettar_HoneycombHubbard_2009}, with the frontier of interesting
sign-problem-free situations still being actively expanded. A key class of 
 applications with direct experimental and theoretical importance 
 are in systems involving ultracold Fermi atoms  \cite{AFQMC_BCS,Hao-2DFG,2DFG_Drut,Bulgac_QMC_BCS-BEC,Trivedi_2011,Alhassid_2013,Alhassid_2014,Scalettar_Tc_2D_Hubbard_2004} and more recently, in cold atom systems with spin-orbit coupling
  \cite{soc-qmc,AFQMC_Rashba_OPLATT}.  
It was also shown that the use of a BCS trial wave function in the unitary Fermi gas
can both reduce projection time (to reach the ground state) and dramatically improve the 
statistical accuracy of the computed ground-state energy
\cite{AFQMC_BCS}. However, with BCS trial wave functions, 
the computation of observables that do not commute 
with the Hamiltonian, and correlation functions, requires additional methodological steps,
which we address in the present paper.

When the sign problem is present, a constraint can be applied 
to remove the exponential increase of the statistical noise. 
The constraint is implemented by introducing a trial wave function,
whose overlap with each random walker is used to define a gauge condition on the sign or phase of that walker.
In such a way the random walks of each walker are constrained to observe the same sign
 \cite{CP-PhysRevB.55.7464,CP-FT-PhysRevLett.83.2777}
or gauge  \cite{phaseless2003}. 
This is an approximation, 
which becomes exact if the trial wave function becomes exact.
%
In a large variety of problems, the constraint has been shown to be a rather loose one 
which yields accurate results and whose accuracy is quite insensitive to the details of the trial wave function \cite{Mario_AFQMC_QC,PhysRevB.88.125132}. 
Typically, 
optimized single Slater determinants (SD) from Hartree-Fock theory or DFT 
are used as trial wave functions. 
In systems where strong pairing is present, it can be expected that the use of a BSC trial 
wave function would improve the results. This was demonstrated in the Fermi gas \cite{AFQMC_BCS}
for the calculation of the ground-state energy. (Indeed  even 
in molecular systems where the electron-electron interaction is repulsive, the 
BCS form can give an improved ansatz \cite{CasulaSorella}, certainly more general than the SD.)
Thus, in constrained-path AFQMC, the implementation of a BCS trial wave function is potentially more
valuable, but the same obstacles in terms of computation of observables must be removed 
as in the sign-problem-free 
cases.

\COMMENTED{
This is still somewhat of a bottle-neck for the methodology, if compared to modern DMC approaches, where highly sophisticated correlated wave functions are routinely used. The difficulty lies in the fact that, while DMC calculations require computation of the value of the wave function together with its derivatives in a given configuration, which can be implemented in a very efficient way, AFQMC requires the value of the overlap between the trial wave function and a given Slater determinant, which, in general, is a computationally formidable task.
In a paper by J. Carlson et al. \cite{AFQMC_BCS}, the very exciting possibility of using a BCS wave function as trial was introduced, and very accurate estimations of the ground state energy of a unitary Fermi gas were obtained. To our knowledge, however, only the calculation of the energy was addressed. The energy is somehow special, since a mixed-estimator is enough to compute it, while other properties require additional non-trivial tools. 
While there exist techniques that exploit the mapping between the repulsive and attractive Hubbard model
through particle-hole transformations, which reinstate a formalism fully relying on Slater determinants,
}

 In this paper we 
 introduce a general approach which removes these obstacles. 
 The approach allows one to explicitly use the BCS wave function 
 as a trial wave function, whether the system is free of or has a sign/phase problem.
 The method we discuss can 
  in principle  be extended to even more general wave functions. Here we focus on lattice models, although, from the methodological point of view, the extension to quantum chemical 
  systems or realistic solids is straightforward.
The BCS wave function, as mentioned, provides a very powerful generalization with respect to single Slater determinants. 
The applicability and accuracy of this approach make it an essential step towards the quantitative description of many strongly-correlated many-body phenomena, including such exotic behaviors as finite-momentum (FFLO) pairing states and high-temperature superconductivity.  

The remainder of the paper is organized as follows. In Section I we introduce the formal definitions of Slater determinants and BCS wave functions, and present some algebraic results that are crucial to the implementation of the methodology.
In Section II we briefly review the AFQMC algorithm and discuss the force bias technique, which allows us to perform a very efficient sampling of auxiliary-field configurations. In Section III, we introduce our new 
technique for computing physical properties. In Section IV, we discuss the 
choice of BCS wave functions and several algorithmic issues.
 Finally, in Section V, we present benchmark results 
 including comparisons with exact diagonalization, 
 before drawing our conclusions in Section VI.

\section{Slater determinants and projected BCS wave functions}
\label{sec:formalism}
Slater determinants and BCS pairing wave functions are essential ingredients in the study of strongly-correlated systems. They serve as the building blocks of several theoretical and computational approaches. Since our purpose is to use BCS wave functions as trial wave functions in AFQMC simulations, we 
first establish our
notation and provide the formal definitions of the wave functions that will be used in the remainder of the paper.

We will limit our attention to the sector of the Hilbert space with fixed number of spin up fermions, $N_{\uparrow}$, and spin down fermions, $N_{\downarrow}$.
We assume $N_{\uparrow} \geq N_{\downarrow}$ for clarity.
The most general Slater determinant under these assumptions has the form:
\begin{equation}
\label{sec1:sd}
| \phi \rangle = \hat{\phi}^{\dagger}_{1, \uparrow}
 \dots  \hat{\phi}^{\dagger}_{N_{\uparrow}, \uparrow} \, \hat{\phi}^{\dagger}_{1, \downarrow} \dots \hat{\phi}^{\dagger}_{N_{\downarrow}, \downarrow}  \, | 0 \rangle
\end{equation}
where:
\begin{equation}
\hat{\phi}^{\dagger}_{i, \sigma} = \sum_{{\bf{r}}} \,  \left(\Phi^{\sigma}\right)_{{\bf{r}}, i} \, \hat{c}^{\dagger}_{{\bf{r}},\sigma}
\end{equation}
with ${\bf{r}}$ denoting lattice sites or basis index, whose dimension will be denoted by $L$.
 The manifold of Slater determinants can thus be parametrized
by $2$ complex $L \times N_{\sigma}$ matrices $\Phi^{\sigma}$, for $\sigma=\uparrow, \downarrow$,
which are made of the orbitals occupied by the independent fermions.

The most general particle-number-projected singlet BCS wave function has the form:
\begin{equation}
\label{sec1:projbcs}
| \, \psi_{BCS} \, \rangle = \left(\prod_{i=1}^{N_{\uparrow} - N_{\downarrow} } \hat{d}^{\dagger}_{i} \right) \,\, 
\frac{1}{N_{\downarrow}!} \left( \sum_{{\bf{r}},{\bf{r}}^{\prime}} F^{}_{{\bf{r}},{\bf{r}}^{\prime}} \, \hat{c}^{\dagger}_{{\bf{r}},\uparrow} \, \hat{c}^{\dagger}_{{\bf{r}}^{\prime},\downarrow} \right)^{N_{\downarrow}} \, |0 \rangle
\end{equation}
It is made of a set of unpaired orbitals:
\begin{equation}
 \hat{d}^{\dagger}_{i}  =  \sum_{{\bf{r}}} \,  \left(D\right)_{{\bf{r}}, i} \, \hat{c}^{\dagger}_{{\bf{r}},\uparrow}
\end{equation}
and a pairing part which describes pairs of fermions in a singlet-state,
with the function $F^{}_{{\bf{r}},{\bf{r}}^{\prime}} $ 
the two-body wave function of the pair. The wave function in \eqref{sec1:projbcs} is parametrized by an $L\times (N_{\uparrow} - N_{\downarrow} )$ complex matrix $D$ and a $L \times L$ complex matrix $F$.
Performing a singular value decomposition,
\begin{equation}
\label{eq:pairing-SVD}
F^{}_{{\bf{r}},{\bf{r}}^{\prime}} = \sum_{\alpha} f_{\alpha} \,\, \mathcal{U}_{{\bf{r}}, \alpha} \, \mathcal{V}^{}_{{\bf{r}}^{\prime},\alpha}
\end{equation}
we 
see that \eqref{sec1:sd} is actually a special case of \eqref{sec1:projbcs} and corresponds to the situation when only $N_{\downarrow}$ of the singular values in $f_{\alpha} $ are non-zero. More generally, \eqref{sec1:projbcs} is a linear combination of Slater determinants.
However, it is much more convenient and computationally efficient compared 
to a generic linear combination, i.e. a multi-determinant, within the AFQMC approach, since the complexity remains comparable to the single determinant case, as we will discuss below. 
Physically, the pairing wave function is clearly the most natural choice 
where fermion pairing correlation is expected.

We next outline some algebraic results for suitable matrix elements of operators between a BCS wave function as in
 \eqref{sec1:projbcs} and a Slater determinant as in \eqref{sec1:sd}.
 Some of the results have been derived before \cite{AFQMC_BCS,Hao-HFB-PhysRevB.95.045144}
 but we include them here to facilitate ensuing discussions.
The central object is the 
 {\it{overlap matrix}}:
\begin{equation}
\mathcal{A} = \left(\left( \Phi^{\uparrow}\right)^{T} D^{\star}  \bigg| \left(\Phi^{\uparrow}\right)^{T} F^{\star} \Phi^{\downarrow}  \right)
\end{equation}
where the vertical bar means that $\mathcal{A} $ is obtained by horizontally stacking
the $N_{\uparrow} \times (N_{\uparrow} - N_{\downarrow} )$ matrix $\left(\Phi^{\uparrow}\right)^{T} D^{\star}$ and the $N_{\uparrow} \times N_{\downarrow}$ matrix $ \left(\Phi^{\uparrow}\right)^{T} F^{\star} \Phi^{\downarrow} $, which results in an $N_{\uparrow} \times N_{\uparrow} $ matrix.
We observe that the actual computation of $\mathcal{A}$, if we store the full matrix $F$ and do not assume any additional property, 
for example translational symmetry, requires three matrix multiplications, with complexity $O(L^2 N)$. 
We will need to calculate the overlap between the BCS wave function and a Slater determinant:
\begin{equation}
S = \left\langle \, \psi_{BCS} \, | \, \phi \right\rangle = (-1)^{N_{\downarrow}(N_{\downarrow}-1)/2}  \,
\det \left(  \mathcal{A} \right) 
\end{equation}
where  the sign factor is due to the convention we use in \eqref{sec1:sd}, writing the Slater determinants with all the spin up first and the spin down later.
The Green function 
matrix elements:
\begin{equation}
\label{eq:green}
\mathcal{G}_{ {\bf{r}} \sigma, {\bf{r}} ^{\prime} \sigma^{\prime}} = \delta_{\sigma,\sigma^{\prime}}  \frac{\langle \, \psi_{BCS} \, |  \hat{c}^{\dagger}_{{\bf{r}},\sigma}
\hat{c}^{}_{{\bf{r}} ^{\prime},\sigma} \, | {\phi} \rangle }{\langle \, \psi_{BCS} \, | \,  {\phi} \rangle } 
\end{equation}
also have simple expressions:
\begin{equation}
{
\mathcal{G}_{ {\bf{r}}  \uparrow, {\bf{r}} ^{\prime}  \uparrow} =  \left( \left(  D^{\star}  \big| \left(F^{\star} \Phi^{\downarrow}  \right)\right) \mathcal{A}^{-1}   \left( \Phi^{\uparrow}\right)^{T} \right)_{ {\bf{r}} ,{\bf{r}} ^{\prime}} }
\end{equation}
and
\begin{equation}
{ 
\mathcal{G}_{{\bf{r}} \downarrow, {\bf{r}} ^{\prime} \downarrow} = 
   \sum_{i = 1}^{N_{\downarrow}}
\left( F^{\dagger} \Phi^{\uparrow} \left( \mathcal{A}^{-1}\right)^{T}\right)_{{\bf{r}}  , N_{u} + i} \left(\Phi^{\downarrow}\right)^{T}_{{i},{\bf{r}} ^{\prime} }}\,, 
\end{equation}
where $N_u = N_{\uparrow} - N_{\downarrow}$ is the number of unpaired orbitals.
We note that the Green functions are $L \times L$ complex matrices and their computation requires inversion of the matrix $\mathcal{A}$ ($\mathcal{O}(N_{\uparrow}^3)$ operations) and matrix multiplications ($\mathcal{O}(L^2 N)$ operations).

Another important component of these simulations is the calculation of two-body correlation 
functions, i.e.,
four point correlators. 
These can be constructed from the Green functions above and
the {\it{anomalous correlators}}:
\begin{equation}
\mathcal{F}_{{\bf{r}}\uparrow,{\bf{r}}^{\prime}\downarrow} = 
 - \sum_{i=1}^{N_{\downarrow}}
\left( \Phi^{\uparrow} \left(\mathcal{A}^{-1}\right)^{T}  \right)_{{\bf{r}},N_{u}+i}  \left(\Phi^{\downarrow}\right)^{T}_{i,{\bf{r}}^{\prime}}
\end{equation}
and
\begin{equation}
\begin{split}
& \overline{\mathcal{F}}_{{\bf{r}}\uparrow,{\bf{r}}^{\prime}\downarrow} = 
 F^{\star}_{{\bf{r}},{\bf{r}}^{\prime}} -  
\left( ( D^{\star} \,  \big| \, F^{\star}\Phi_{\downarrow} )
\mathcal{A}^{-1}  \left( \left(\Phi^{\uparrow}\right)^{T} F^{\star} \right)\right)_{{\bf{r}},{\bf{r}}^{\prime}}\,. 
\end{split}
\end{equation}
These computations share the same complexity as the Green functions,
and have similar definitions.
For example, $\mathcal{F}_{{\bf{r}},{\bf{r}}^{\prime}}$ gives 
the matrix element of the operator $\hat{c}^{}_{{\bf{r}},\uparrow} \hat{c}^{}_{{\bf{r}}^{\prime},\downarrow} $ between the Slater determinant $|\phi\rangle$ 
and a BCS wave function which does {\it{not}} conserve the number of particles, but which is defined by the same pairing matrix $F$ and the same set of unpaired orbitals (i.e., 
the parent wave function from which the $\psi_{BCS}$ is derived via number-projection).
With some care we can apply Wick's theorem to 
obtain the expressions for the two-body correlations.
For example,
\begin{equation}
\mathcal{C}_{{\bf{r}}_1 \downarrow,{\bf{r}}_2 \downarrow,{\bf{r}}_3 \uparrow,{\bf{r}}_4 \uparrow}
= \frac{\langle \, \psi_{BCS} \, | 
\hat{c}^{\dagger}_{{\bf{r}}_1,\downarrow}
\hat{c}^{}_{{\bf{r}}_2,\downarrow}  \hat{c}^{\dagger}_{{\bf{r}}_3,\uparrow}
\hat{c}^{}_{{\bf{r}}_4,\uparrow} \, | {\phi} \rangle }{\langle \, \psi_{BCS} \, | \,  {\phi} \rangle }  
\end{equation}
is given by 
\begin{equation}
\begin{split}
& \mathcal{C}_{{\bf{r}}_1 \downarrow,{\bf{r}}_2 \downarrow,{\bf{r}}_3 \uparrow,{\bf{r}}_4 \uparrow}
= \mathcal{G}_{{\bf{r}_1} \downarrow, {\bf{r}}_2 \downarrow}  \mathcal{G}_{{\bf{r}_3} \uparrow, {\bf{r}}_4 \uparrow} \\
& - \overline{\mathcal{F}}_{{\bf{r}}_3 \uparrow,{\bf{r}}_1 \downarrow} \mathcal{F}_{{\bf{r}}_4 \uparrow,{\bf{r}}_2 \downarrow}\,,
\end{split}
\end{equation}
and 
\begin{equation}
{
\mathcal{C}_{{\bf{r}}_1 \sigma, {\bf{r}}_2  \sigma, {\bf{r}}_3 \sigma,  {\bf{r}}_4 \sigma}
= \frac{\langle \, \psi_{BCS} \, | 
\hat{c}^{\dagger}_{{\bf{r}}_1,\sigma}
\hat{c}^{}_{{\bf{r}}_2,\sigma}  \hat{c}^{\dagger}_{{\bf{r}}_3,\sigma}
\hat{c}^{}_{{\bf{r}}_4,\sigma} \, | {\phi} \rangle }{\langle \, \psi_{BCS} \, | \,  {\phi} \rangle }  
}
\end{equation}
is given by 
\begin{equation}
\begin{split}
&
\mathcal{C}_{ {\bf{r}}_1 \sigma, {\bf{r}}_2  \sigma, {\bf{r}}_3 \sigma,  {\bf{r}}_4 \sigma}
=   \delta_{{\bf{r}}_3, {\bf{r}}_2} 
 \mathcal{G}_{{\bf{r}}_1  \sigma,  {\bf{r}}_4 \sigma}   \\
& + \mathcal{G}_{ {\bf{r}}_1 \sigma, {\bf{r}}_2 \sigma}
\mathcal{G}_{{\bf{r}}_3  \sigma, {\bf{r}}_4  \sigma}
-  \mathcal{G}_{ {\bf{r}}_1 \sigma, {\bf{r}}_4  \sigma}
 \mathcal{G}_{ {\bf{r}}_3 \sigma, {\bf{r}}_2 \sigma}
\end{split}
\end{equation}

\section{The AFQMC algorithm with force bias}
\label{sec:method}
With the formalism introduced in the previous section, 
we now describe the technique. We will focus on lattice models
and present the algorithm in the constrained-path formulation, commonly referred to as constrained-path
auxiliary-field quantum Monte Carlo (CP-AFQMC).
However, as mentioned, the algorithm we introduce can be generalized to the phaseless AFQMC
approach for molecules and solids. It is also straightforward to apply it to standard determinantal Monte Carlo.

For the purpose of illustration and concreteness, we 
start with a single-band Hubbard model:
\begin{equation}
\hat{H} = \hat{K} + \hat{V}
\end{equation}
with:
\begin{equation}
 \hat{K} = -t \sum_{\langle {\bf{r}} \, , \, {\bf{r}^{\prime}} \rangle \, \sigma }    \hat{c}^{\dagger}_{{\bf{r}},\sigma} 
\hat{c}^{}_{{\bf{r}^{\prime}},\sigma} + h. c. , \quad \hat{V} = U \sum_{{\bf{r}} } \hat{n}_{{\bf{r}},\uparrow} \hat{n}_{{\bf{r}},\downarrow} 
\end{equation}
where, 
as usual, $\hat{n}_{{\bf{r}},\sigma} = \hat{c}^{\dagger}_{{\bf{r}},\sigma} 
\hat{c}^{}_{{\bf{r}},\sigma}  $ and the brackets denote nearest-neighbor sites. 
The approach relies on the imaginary time evolution operator $\exp({-\tau \hat{H}})$
which, for large $\tau$, projects onto the ground state wave function $\left| \, \Psi_0 \, \right\rangle$ of the model. For any chosen initial wave function $\left| \, \phi_0 \, \right\rangle$, 
not orthogonal to the ground state, we have: 
\begin{equation}
\label{projection}
\lim_{\tau \to +\infty} e^{-\tau \left(\hat{H} - E_0\right)} \, \left| \, \phi_0 \, \right\rangle \propto \left| \, \Psi_0 \, \right\rangle\,.
\end{equation}
For simplicity, we assume $ \left| \, \phi_0 \, \right\rangle $ to be a single Slater determinant, for example, the non-interacting Fermi sea. 
(It is straightforward to use a linear combination of Slater determinants. 
Moreover, it may become important to use  a BCS initial wave function, especially
in the path-integral formalism typically used for sign-problem-free situations. This will be 
discussed in the next section.)

The relation \eqref{projection}  defines a path integral 
in the Hilbert space of the system which connects the initial wave function to the ground state wave function. The essence of CP-AFQMC is to map \eqref{projection} onto a random walk in the manifold of Slater determinants, which can be efficiently sampled through Monte Carlo methods.
The key tool to achieve this is a combination of the Trotter decomposition and the Hubbard-Stratonovich (HS) transformation:
\begin{equation}
\label{hs}
e^{-\tau \left(\hat{H} - E_0\right)}  \simeq \left(  e^{-\delta \tau \left(\hat{H} - E_0\right)}\right)^n 
\simeq \left( \int d{\bf{x}} \, p({\bf{x}}) \, \hat{B}({\bf{x}}) \right)^n 
\end{equation}
In \eqref{hs} the integration is carried over the configurations ${\bf{x}}$ of an auxiliary-field defined on the lattice, each configuration being weighted by a probability density $p({\bf{x}})$. The operator $\hat{B}({\bf{x}})$ is a one-body propagator, describing the evolution of a non-interacting Fermi gas embedded in an external random field. The crucial advantage of using the HS transformation is that the operator $ \hat{B}({\bf{x}}) $ transforms Slater determinants into Slater determinants: whenever $| \phi \rangle$ is a Slater determinant, $ \hat{B}({\bf{x}}) \, | \phi \rangle$ is also a Slater determinant.
More explicitly, for the attractive Hubbard model, the auxiliary-field is an Ising field on the lattice:
${\bf{x}}({\bf{r}}) = \pm 1$, with uniform probability density $p({\bf{x}}) = 1/2^L$, and the propagator $\hat{B}({\bf{x}})$ can be  constructed using the operator identity:
\begin{equation}
\label{eq:HS-ising}
e^{-\delta\tau U \hat{n}_{{\bf{r}},\uparrow} \hat{n}_{{\bf{r}},\downarrow}  } 
=e^{-\delta\tau U (\hat{n}_{{\bf{r}}} -  1)/2 } \sum_{x=\pm 1} \frac{1}{2} e^{\gamma x (\hat{n}_{{\bf{r}}} -  1)}
\end{equation}
where $\hat{n}_{{\bf{r}}} = \hat{n}_{{\bf{r}},\uparrow} + \hat{n}_{{\bf{r}},\downarrow}$ is the local particle density operator.
Note that there are variants of this form which can affect the Monte Carlo sampling efficiency
and the systematic accuracy of the constraints (see, e.g., Ref.~\cite{PhysRevB.88.125132}), but 
we will not distinguish them here and will focus on the general formalism instead.
The one body propagator takes the form:
\begin{equation}
\label{operatorB}
\hat{B}({\bf{x}}) = e^{\tau E_0} \, e^{-\delta\tau\hat{K}/2} \, \prod_{{\bf{r}}} \hat{b}_{{\bf{r}}}(x({\bf{r}}))\,e^{-\delta\tau\hat{K}/2} 
\end{equation}
with:
\begin{equation}
 \hat{b}_{{\bf{r}}}(x) = e^{-\delta\tau U (\hat{n}_{{\bf{r}}} -  1)/2 } \, e^{\gamma x (\hat{n}_{{\bf{r}}} -  1)}
\end{equation}

Using these ingredients, 
Eq.~\eqref{projection} is mapped onto an ensemble of paths in the manifold of Slater determinants and the ensemble average recovers the fully correlated problem. This average corresponds to the multidimensional integral over the space-time dependent auxiliary field configurations in \eqref{hs}. In order to compute this integral, an importance sampling scheme 
is implemented through the introduction of a trial wave function $ \left| \, \psi_T \, \right\rangle$, which is assumed to be a good approximation to the ground state wave function. At imaginary time
 $\tau = n \delta \tau$, a stochastic linear combination of the form:
\begin{equation}
\label{sec2:stochlincomb}
\left| \, \Psi(n) \, \right\rangle = \sum_{w} \mathcal{W}_{w}(n) \frac{| \phi^{w}(n) \rangle}{\langle \psi_T \, | \phi^{w}(n) \rangle}
\end{equation}
is generated, where 
$| \phi^{w}(n) \rangle$
are Slater determinants, i.e., the {\it{walkers}}, which are labeled by $w$, and
$\mathcal{W}_{w}(n)$ are their weights. 

At the initial time $n=0$, we let $| \phi^{w}(n=0) \rangle = | \phi_0 \rangle$, making all the walkers start from the  initial wave function, and the weights are $\mathcal{W}_{w}(n=0) = 1 $, such that $\left| \, \Psi(n) \, \right\rangle = | \phi_0 \rangle$, apart from an irrelevant normalization factor. 

For $n > 0$, to build $\left| \, \Psi(n+1) \, \right\rangle$ starting from $\left| \, \Psi(n) \, \right\rangle$, we proceed as follows. For each walker $w$, to apply the kinetic energy
(or one-body part of the Hamiltonian) term:
\begin{equation}
e^{-\delta\tau\hat{K}/2} \, \mathcal{W}_{w}(n) \frac{| \phi^{w}(n) \rangle}{\langle \psi_T \, | \phi^{w}(n) \rangle}\,,
\end{equation}
we have 
\begin{equation}
\mathcal{W}_{w}  \frac{\langle \psi_T \, | e^{-\delta\tau\hat{K}/2} \phi^{w} \rangle}{\langle \psi_T \, | \phi^{w} \rangle} \, \frac{e^{-\delta\tau\hat{K}/2} | \phi^{w} \rangle}{\langle \psi_T \, | e^{-\delta\tau\hat{K}/2} \phi^{w} \rangle}\,,
\end{equation}
where we have omitted 
the dependence on $n$ to keep the notations simple. This implies that the kinetic contribution to the propagation in imaginary time leads to the simple updates:
\begin{equation}
\mathcal{W}_{w} \to    \mathcal{W}_{w} \, \frac{\langle \psi_T \, | e^{-\delta\tau\hat{K}/2} \phi^{w} \rangle}{\langle \psi_T \, | \phi^{w} \rangle}
\end{equation}
and
\begin{equation}
| \phi^{w} \rangle \to e^{-\delta\tau\hat{K}/2} |\phi^{w} \rangle\,.
\end{equation}

Let us now turn to the 
interaction part:
\begin{equation}
\left( \int d{\bf{x}} \, p({\bf{x}}) \,  \prod_{{\bf{r}}} \hat{b}_{{\bf{r}}}(x({\bf{r}}))\, \right) \, \mathcal{W}_{w} \frac{| \phi^{w} \rangle}{\langle \psi_T \, | \phi^{w} \rangle}\,,
\end{equation}
which 
can be rewritten as:
\begin{equation}
\label{eq:ImpSamp}
\int d{\bf{x}}  \, \pi({\bf{x}})  \mathcal{W}_{w} \, \,
 \frac{ \prod_{{\bf{r}}} \hat{b}_{{\bf{r}}}(x({\bf{r}}))\,\,| \phi^{w} \rangle}{\langle \psi_T \, |  \prod_{{\bf{r}}} \hat{b}_{{\bf{r}}}(x({\bf{r}}))\, \phi^{w} \rangle}\,,
\end{equation}
where
\begin{equation}
\pi({\bf{x}}) = p({\bf{x}}) \frac{\langle \psi_T \, | \prod_{{\bf{r}}} \hat{b}_{{\bf{r}}}(x({\bf{r}})) \, | \, \phi^{w} \rangle}{\langle \psi_T \, | \phi^{w} \rangle}\,. 
\end{equation}
The idea of importance sampling 
is to sample ${\bf{x}}$ not from 
the bare $p({\bf{x}})$, but according to a suitable approximation for the function $\pi({\bf{x}})$, 
which 
favors walkers with larger overlap with the trial wave function. 
Such importance sampling can improve the efficiency dramatically even with a modest 
trial wave function, because of the large dimensionality involved (i.e., of ${\bf x}$ and 
the fact that this is repeatedly applied over many iterations $n$).
\COMMENTED{
If the trial wave function is a good approximation to the ground state wave function, such a sampling is naturally desired, and turns out to be essential to the success of QMC simulations of strongly correlated systems.
}

In the present paper we implement importance sampling in the above by using a force bias 
 \cite{phaseless2003}.
We introduce 
an additional (not normalized) probability density $\tilde{p}({\bf{x}})$, with $\tilde{\mathcal{N}} = \int d{\bf{x}} \, \tilde{p}({\bf{x}})$ and 
rewrite Eq.~(\ref{eq:ImpSamp}) as
\begin{equation}
\int d{\bf{x}} \, \frac{\tilde{p}({\bf{x}})}{\tilde{\mathcal{N}}} \, \frac{  \tilde{\mathcal{N}} \, \pi({\bf{x}}) }{\tilde{p}({\bf{x}})} \,  \mathcal{W}_{w} \, \,
 \frac{ \prod_{{\bf{r}}} \hat{b}_{{\bf{r}}}(x({\bf{r}}))\,\,| \phi^{w} \rangle}{\langle \psi_T \, |  \prod_{{\bf{r}}} \hat{b}_{{\bf{r}}}(x({\bf{r}}))\, \phi^{w} \rangle}\,,
\end{equation}
and choose:
\begin{equation}
\tilde{p}({\bf{x}}) =  \prod_{{\bf{r}}}  \left( 1 + \gamma x({\bf{r}}) \frac{ \langle \psi_T \, | ( \hat{n}_{{\bf{r}}} - 1 ) \, |
\phi^{w} \rangle}{ \langle \psi_T \, | \phi^{w} \rangle} \right)\,.
\end{equation} 
The force bias above is a discrete version of the typical form in a shifted Gaussian probability 
 \cite{phaseless2003}, and allows us to continue to use the Ising auxiliary-fields from Eq.~(\ref{eq:HS-ising}). It is closely related to the form  used in Ref.~\cite{Hao-2DFG},
 obtained from $\pi({\bf{x}})$ by expanding the exponential up to order $\sqrt{\delta\tau}$, as an approximation for $\pi({\bf{x}})$ in the small time step limit. 

In summary, to implement the interaction part of the propagation, we first compute
the mixed estimator of the density:
\begin{equation}
\tilde{n}({\bf{r}}) =  \frac{ \langle \psi_T \, |  \hat{n}_{{\bf{r}}}  \, |
\phi^{w} \rangle}{ \langle \psi_T \, | \phi^{w} \rangle}\,,
\end{equation}
sample a configuration ${\bf{x}}$ drawn from $\tilde{p}({\bf{x}})$, 
which consists of sampling
independent random variables on each lattice site, and update:
\begin{equation}
\mathcal{W}_{w} \to \mathcal{W}_{w}  \, \frac{  \tilde{\mathcal{N}} \, \pi({\bf{x}}) }{\tilde{p}({\bf{x}})} 
\end{equation}
and
\begin{equation}
| \phi^{w} \rangle \to  \prod_{{\bf{r}}} \hat{b}_{{\bf{r}}}(x({\bf{r}})) \, | \, \phi^{w} \rangle\,.
\end{equation}
Iterating this procedure and extrapolating to infinite total imaginary time, infinite number of walkers, and zero time step, the stochastic linear combination $\left| \, \Psi(n) \, \right\rangle $
converges to the ground state wave function. Note that at the limit of $\delta\tau \rightarrow 0$, 
the constraint is automatically imposed by the importance sampling in the case of a sign problem.
With a twist boundary condition or when a magnetic field is imposed, a ``weak'' phase problem arises 
for which a straightforward 
generalization can be applied \cite{Chiachen-simple-phase-problem}. When a more intrinsic
phase problem (with realistic electron-electron interactions, for example) is present, a phaseless 
approximation needs to be imposed which requires an additional projection beyond the 
importance sampling  \cite{phaseless2003}.

\COMMENTED{
However, in order to control the fermion sign problem, the random walk is
constrained by the condition:
\begin{equation}
\langle \psi_T \, | \phi^{w}(n) \rangle > 0
\label{eqn:nodal_surface}
\end{equation}
This is an approximation, which is exact if $\psi_T$ coincides with the actual 
ground state wave function or (with some caveat) if there is no sign problem.
}

\section{Computation of observables}

Suppose we choose a BCS wave function, $|\psi_{BCS}\rangle$, as the trial wave function for a CP-AFQMC calculation, or use $|\psi_{BCS}\rangle$ as the initial wave wavefunction $|\phi_0\rangle$ 
in a path-integral formulation as is adopted in more standard sign-problem-free calculations.  
To compute a general expectation value of a physical observable in either case,
the propagation of  $|\psi_{BCS}\rangle$ by the one-body propagator in 
Eq.~(\ref{operatorB}) is necessary. This causes a computational difficulty as we discuss below, 
followed by a proposed solution.

The simplest kind of estimator that we can build is a {\it{mixed estimator}}, defined as:
\begin{equation}
\label{eq:mixed-est}
\mathcal{O}_{mixed} = \frac{\langle  \, \psi_{BCS} | \, \hat{O} \, | \, \Psi_0 \rangle}{\left\langle \psi_{BCS} \, |  \, \Psi_0 \right\rangle}
\end{equation}
Using the stochastic linear combination \eqref{sec2:stochlincomb} as an approximation 
to the ground state $| \, \Psi_0 \rangle$ we get immediately:
\begin{equation}
\mathcal{O}_{mixed} = 
\frac{\sum_{w} \mathcal{W}_{w} \frac{\langle  \, \psi_{BCS} | \, \hat{O} \, | \phi^{w} \rangle}{\vphantom{\tilde{\psi}}\langle \psi_{BCS} \, | \phi^{w} \rangle}}
{\sum_{w} \mathcal{W}_{w}}
\end{equation}
This estimator can be readily computed using the sampling scheme described above, 
combined with the algebraic relations that we introduced in Sec.~\ref{sec:formalism}.
This is the approach used to compute the Bertsch parameter in Ref.~\cite{AFQMC_BCS}
and the equation of state in the two-dimensional Fermi gas \cite{Hao-2DFG}.

For a general observable, or correlation functions, the mixed estimator is insufficient 
and we need to compute the {\it{pure estimator}}:  
\COMMENTED{
In the case when the operator $\hat{O}$ commutes with the Hamiltonian
operator, and thus with the imaginary time propagator $\exp(-\tau \hat{H})$, the mixed estimator coincides with the {\it{pure estimator}}:
}
\begin{equation}
\label{eq:pure-est}
\mathcal{O}_{pure} =  \frac{\langle  \, \Psi_{0} | \, \hat{O} \, | \, \Psi_0 \rangle}{\left\langle \Psi_{0} \, |  \, \Psi_0 \right\rangle}\,.
\end{equation}
In the special case when the operator $\hat{O}$ commutes with the Hamiltonian
operator, and thus with the imaginary time propagator $\exp(-\tau \hat{H})$, the mixed estimator 
in Eq.~(\ref{eq:mixed-est}) becomes equivalent to 
the pure estimator in Eq.~(\ref{eq:pure-est}). In general, however,
the two are different and the mixed estimator is biased. 

We introduce here a 
new technique to compute the pure expectation value of Eq.~(\ref{eq:pure-est}) 
within the auxiliary-field framework when 
a BCS wave function is involved. Below we describe the technique 
in a CP-AFQMC calculation. It is straightforward to  generalize it to the path-integral formulation 
for sign-problem-free calculations, or indeed in several other contexts (for example mean-field  
calculations by projection), which we will discuss in Sec.~\ref{ssec:extensions}.

We make use of the following:
\begin{equation}
\langle  \, \Psi_{0} |  \simeq \langle  \, \psi_{BCS} | \, e^{-\overline{\tau} (\hat{H} - E_0)}
\end{equation}
for large 
$\overline{\tau} = m \delta\tau$, 
which will be accomplished by carrying out many discrete time-steps.
We first consider, for the purpose of illustration, a single time-step:
\begin{equation}
\label{sec2:pure}
\begin{split}
& \langle  \, \psi_{BCS} | \, e^{-\delta{\tau} (\hat{H} - E_0)} | \, \sum_{w} \mathcal{W}_{w}(n)  \frac{  \hat{O} \, | \phi^{w} (n) \rangle }{\langle \psi_{BCS} \, | \phi^{w} (n) \rangle} \\
& \simeq   \int d{\bf{x}} \, p({\bf{x}}) \, \sum_{w} \mathcal{W}_{w}(n) 
\frac{\langle \psi_{BCS} | \, \hat{B}({\bf{x}}) \, \hat{O} \, | \phi^{w} (n) \rangle }{\langle \psi_{BCS} \, | \phi^{w} (n) \rangle} \\
& =   \int d{\bf{x}} \, \pi({\bf{x}}) \, \sum_{w} \mathcal{W}_{w}(n) 
\frac{\langle \psi_{BCS} | \, \hat{B}({\bf{x}}) \, \hat{O} \, | \phi^{w} (n) \rangle }{\langle \psi_{BCS} \, |  \hat{B}({\bf{x}}) \phi^{w} (n) \rangle} 
\end{split}
\end{equation}
where we have reinstated the {\it{time}} label $n$ which is useful here and we have used the same manipulations described in the previous section.

In the usual approach to \eqref{sec2:pure}, 
referred to as {\it{back propagation}}  \cite{CP-PhysRevB.55.7464,BP-mol-Mario}, a walker  
$|\phi^{w} (n) \rangle$ is first propagated in the forward direction as described in the previous 
section, and the
path 
${\bf{x}}$ that is
 sampled is then used to  explicitly transform:
\begin{equation}
\label{sec2:leftupdate}
\langle \psi_{BCS} |  \to \langle \psi_{BCS} | \, \hat{B}({\bf{x}}) 
\end{equation}
building a new bra which is used in combination with $|\phi^{w} (n) \rangle$ 
to compute the estimator. The update \eqref{sec2:leftupdate}
can be implemented for a BCS wave function, resulting in another BCS wave function with
a different set of unpaired orbitals and a different pairing matrix. 
More specifically 
in the expression in Eq.~(\ref{eq:pairing-SVD}), the unpaired orbitals and 
$\mathcal{U}_{{\bf{r}}, \alpha}$ are multiplied by the spin-$\uparrow$ part of the propagator, 
while $\mathcal{V}^{}_{{\bf{r}}^{\prime},\alpha}$ is multiplied by the spin-$\downarrow$ part of the 
propagator, similar to how the $\uparrow$- and $\downarrow$-spin parts of a single Slater
determinant are propagated. However, when iterated many times (to reach large $m$), 
the orbitals will become numerically contaminated. With Slater determinants, this can 
be controlled by  Gram-Schmidt re-orthonormalization, in which all but the 
resulting orthonormal orbitals can be discarded without affecting the algorithm. With $\langle \psi_{BCS}|$, this is not the case, and we did not find a satisfactory way to keep an arbitrarily 
high level of numerical stability.



Guided by the desire to have more general and numerically stable calculations 
in the BCS case, we devise a different approach, which we present in the remainder of this section.
Let us consider the case:
\begin{equation}
\hat{O} = \hat{c}^{\dagger}_{\mu,\sigma}  \hat{c}^{}_{\nu,\sigma} 
\end{equation}
with $| \mu \rangle = \sum_{{\bf{r}}} \langle {\bf{r}} |  \mu \rangle  | {\bf{r}} \rangle$
 and similarly for $| \nu \rangle $. 
We use the operator identity:
\begin{equation}
\hat{B}({\bf{x}}) \,  \hat{c}^{\dagger}_{\mu,\sigma}  \hat{c}^{}_{\nu,\sigma} 
=  \hat{c}^{\dagger}_{\hat{B}({\bf{x}}) \mu,\sigma}  \hat{c}^{}_{(\hat{B}({\bf{x}})^{-1})^{\dagger}\nu,\sigma} \, \hat{B}({\bf{x}})
\end{equation}
which rests on the fact that $\hat{B}({\bf{x}})$ is a single particle propagator. Intuitively, if we
want $\hat{B}({\bf{x}})$ to ``jump'' over a $\hat{c}^\dagger_{\mu}$ or $\hat{c}_\nu$
operator, 
we need to let the orbitals $| \mu \rangle$ 
and $| \nu \rangle $ evolve under the action of $\hat{B}({\bf{x}})$, in the forward and backward time direction, respectively:
\begin{equation}
\label{sec2:stories}
| \mu^{\prime} \rangle = \hat{B}({\bf{x}}) \, | \mu \rangle, \quad
| \nu^{\prime} \rangle = (\hat{B}({\bf{x}})^{-1})^{\dagger} \, | \nu \rangle
\end{equation}
We can thus write:
\begin{equation}
\label{sec2:frompuretomixed}
\begin{split}
& \frac{\langle \psi_{BCS} | \, \hat{B}({\bf{x}}) \, \hat{c}^{\dagger}_{\mu,\sigma}  \hat{c}^{}_{\nu,\sigma}  \, | \phi^{w} (n) \rangle }{\langle \psi_{BCS} \, |  \hat{B}({\bf{x}}) \phi^{w} (n) \rangle} \\
& = \frac{\langle \psi_{BCS} | \,  \, \hat{c}^{\dagger}_{\mu^{\prime},\sigma}  \hat{c}^{}_{\nu^{\prime},\sigma} \, \hat{B}({\bf{x}}) \, | \phi^{w} (n) \rangle }{\langle \psi_{BCS} \, |  \hat{B}({\bf{x}}) \phi^{w} (n) \rangle}
\end{split}
\end{equation}
The right-hand side 
of \eqref{sec2:frompuretomixed} closely resembles the usual mixed estimator
and has the further advantage that we can apply Gram-Schmidt decomposition to the Slater determinant as it is propagated:
\begin{equation}
| \phi^{\prime} \rangle = \hat{B}({\bf{x}}) \, | \phi^{w} (n)  \rangle\,,
\end{equation}
without changing the result. The difference is that we need to follow the evolution of the orbitals \eqref{sec2:stories}, which is straightforward to implement. 

The generalization of the above  single time-step procedure  
 to more time steps $m > 1$ follows immediately by iteration. 
In particular, if one wishes to compute the full one-body density matrix, it is sufficient to compute the evolution matrix:
\begin{equation}
\label{sec2:Bstory}
\mathcal{B}_{{\bf{r}},{\bf{r}}^{\prime}}(m) = \langle {\bf{r}} | \hat{B}({\bf{x}}_1) \dots \hat{B}({\bf{x}}_m)  \, | {\bf{r}}^{\prime} \rangle
\end{equation}
as well as the corresponding one for the backward evolution
involving the product of the inverse adjoint of the propagator matrices.

We observe that, although we are able to stabilize the Slater determinant as it is propagated forward,
we still have to deal with matrices of the form in Eq.~\eqref{sec2:Bstory}, which unavoidably leads to numerical instabilities. To control these instabilities, we assume that:
\begin{equation}
| \phi^{\prime} \rangle = \hat{\phi^{\prime}}_{1,\uparrow}^{\dagger} \,  \hat{\phi^{\prime}}_{2,\uparrow}^{\dagger} \dots |0 \rangle
\end{equation}
is composed of orthonormal orbitals, which can be achieved easily via the forward stabilization already in place in AFQMC. 
We use the following properties of creation and destruction operators:
\begin{equation}
\label{sec2:tricks}
\begin{split}
& \hat{c}^{\dagger}_{\mu^{\prime},\sigma}  \hat{c}^{}_{\nu^{\prime},\sigma} +
 \hat{c}^{}_{\nu^{\prime},\sigma} \hat{c}^{\dagger}_{\mu^{\prime},\sigma}  = \langle \nu^{\prime} | \mu^{\prime} \rangle \\
& \hat{c}^{}_{\nu^{\prime},\sigma} \, | \phi^{\prime} \rangle  =   \hat{c}^{}_{\tilde{\nu}^{\prime},\sigma} \, | \phi^{\prime} \rangle \\
 & \hat{c}^{\dagger}_{\mu^{\prime},\sigma}  \, | \phi^{\prime} \rangle  =  \hat{c}^{\dagger}_{\tilde{\mu}^{\prime},\sigma}  | \phi^{\prime} \rangle\,, 
\end{split}
\end{equation}
where:
\begin{equation}
\label{sec2:spac}
\begin{split}
& | \tilde{\nu}^{\prime} \rangle = \sum_{i=1}^{N_{\sigma}} \langle \phi^{\prime}_{i,\sigma} | \nu^{\prime} \rangle \, 
| \phi^{\prime}_{i,\sigma} \rangle \\
& | \tilde{\mu}^{\prime} \rangle = | {\mu}^{\prime}  \rangle -  \sum_{i=1}^{N_{\sigma}} \langle \phi^{\prime}_{i,\sigma} | \mu^{\prime} \rangle \, 
| \phi^{\prime}_{i,\sigma} \rangle\,.
\end{split}
\end{equation}
In \eqref{sec2:tricks}, the first equality follows from canonical anticommutation relations, while the other relations are closely related to the Pauli exclusion principle. Intuitively, when we destroy an orbital
in $| \phi'\rangle$, only the component in the linear span of the orthonormal orbitals defining $ | \phi^{\prime} \rangle $ 
contributes;
for the  creation operator, 
the ``opposite'' is true, where we have to project out all linear dependencies \cite{ettoreGAP}.

We thus obtain for the Green function: 
\begin{equation}
\begin{split}
&
\hat{c}^{\dagger}_{\mu^{\prime},\sigma}  \hat{c}^{}_{\nu^{\prime},\sigma} \, | \phi^{\prime} \rangle 
= \hat{c}^{\dagger}_{\mu^{\prime},\sigma}  \hat{c}^{}_{\tilde{\nu}^{\prime},\sigma} \, | \phi^{\prime} \rangle \\
&=  \langle \tilde{\nu}^{\prime}  | \mu^{\prime} \rangle \, | \phi^{\prime} \rangle - \hat{c}^{}_{\tilde{\nu}^{\prime},\sigma} \,  \hat{c}^{\dagger}_{\tilde{\mu}^{\prime},\sigma}  \, \, | \phi^{\prime} \rangle \\
& = \langle \tilde{\nu}^{\prime}  | \mu^{\prime} \rangle \, | \phi^{\prime} \rangle 
+  \hat{c}^{\dagger}_{\tilde{\mu}^{\prime},\sigma}  \,  \hat{c}^{}_{\tilde{\nu}^{\prime},\sigma} \,\, | \phi^{\prime} \rangle
\end{split}
\end{equation}
where, in the last line, we have used the fact that $\langle \tilde{\nu}^{\prime}  |   \tilde{\mu}^{\prime} \rangle = 0$ by construction.
In this way, instead of \eqref{sec2:stories}, we have the modified evolution:
\begin{equation}
| \mu \rangle \to | \tilde{\mu}^{\prime} \rangle, \quad | \nu \rangle \to | \tilde{\nu}^{\prime} \rangle
\end{equation}
and its obvious iteration for $m>1$.
The key advantage is that, thanks to \eqref{sec2:spac}, the projection on the linear span of the orbitals defining $| \phi^{\prime} \rangle$ and on its orthogonal complement, we can keep the instability under control. Empirically, we always find that we can achieve a robust extrapolation to $m \to +\infty$, while keeping the statistical noise level small enough.
This completes 
the full CP-AFQMC technique which uses a BCS wave function as a trial wave function and enables the computation of the ground-state expectation of any physical property. 

\COMMENTED{Before showing some results and benchmarks, we find it useful to discuss the problem of how to select the actual trial wave function  \eqref{sec1:projbcs}, that is how to find the optimal unpaired orbitals and pairing matrix $F$ for a given model.}

\section{Results}

\begin{figure*}[ht]
\begin{center}
\includegraphics[width=15.0cm, angle=0]{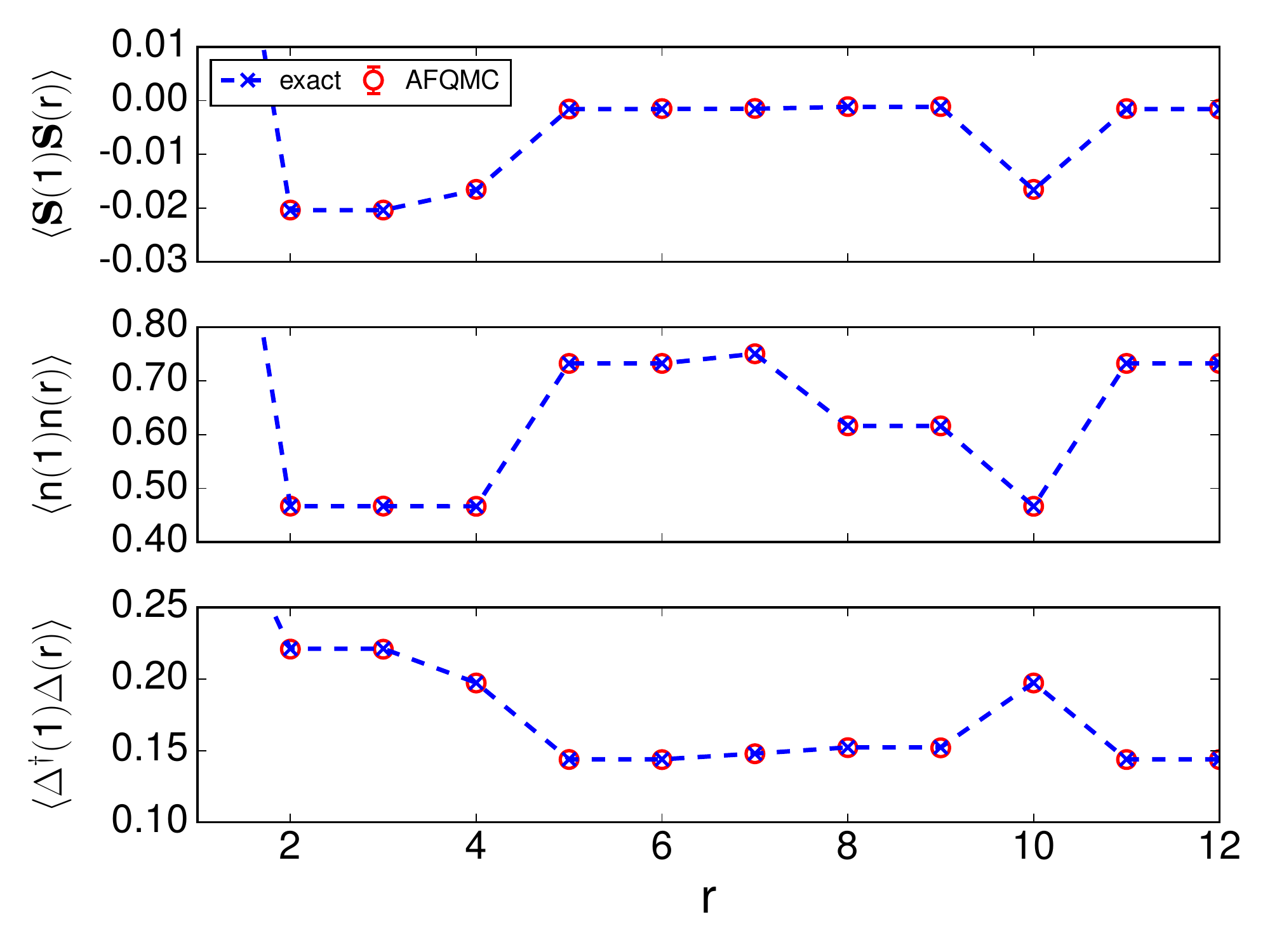}
\caption{(Color online) Comparison  of AFQMC with BCS trial wave functions with exact results. 
The system is a $3 \times 4$  lattice under periodic boundary conditions, with 
 $N_{\uparrow} = N_{\downarrow} = 5$ fermions and an interaction strength $U/t = -8$. Each correlation function is plotted versus site number.}
\label{fig:spin-balanced}
\end{center}
\end{figure*}

In this section we present two sets of benchmark results obtained using the method we have described. We compare these results with exact diagonalization (ED) calculations in small 
lattices and standard CP-AFQMC with Slater determinants in larger lattices.

\subsection{Spin-balanced system}
As a demonstration of the accuracy of the methodology, we have performed several benchmark calculations, which we compare with ED
for a system of $N_{\uparrow} = N_{\downarrow} = 5$ fermions moving on a two-dimensional
lattice with $3 \times 4$ sites in periodic boundary conditions and an interaction strength $U/t = -8$.
In Fig.~\ref{fig:spin-balanced}, we show the comparison for three different correlations functions,
which are important 
for the study of correlated systems. The
spin correlation function $\left\langle {\bf{\hat{S}}}({\bf{r}}) \cdot  {\bf{\hat{S}}}({\bf{r}^{\prime}}) \right\rangle$,
the density correlation function $\left\langle {\hat{n}}({\bf{r}})   {\hat{n}}({\bf{r}^{\prime}}) \right\rangle$, 
and the on-site s-wave pairing correlation function
$\left\langle {\hat{\Delta}}^{\dagger}({\bf{r}})   {\hat{\Delta}}({\bf{r}^{\prime}}) \right\rangle$.
The 
spin operator is defined as
\begin{equation}
{\bf{\hat{S}}}({\bf{r}}) = \frac{1}{2} \sum_{\sigma,\sigma^{\prime}} \vec{\sigma}_{\sigma,\sigma^{\prime}}  \, 
\hat{c}^{\dagger}_{{\bf{r}},\sigma} \hat{c}^{}_{{\bf{r}},\sigma^{\prime}} \,,
\end{equation}
with $\vec{\sigma}_{\sigma,\sigma^{\prime}}  $ denoting the elements of the Pauli matrices.
The density operator is ${\hat{n}}({\bf{r}})  = \sum_{\sigma} \hat{c}^{\dagger}_{{\bf{r}},\sigma} \hat{c}^{}_{{\bf{r}},\sigma} $,
while the pairing operator is 
${\hat{\Delta}}^{}({\bf{r}}) =  \hat{c}^{}_{{\bf{r}},\downarrow}   \hat{c}^{}_{{\bf{r}},\uparrow}$.
\COMMENTED{
\begin{equation}
{\hat{\Delta}}^{}({\bf{r}}) =  \hat{c}^{}_{{\bf{r}},\downarrow}   \hat{c}^{}_{{\bf{r}},\uparrow} 
\end{equation}
}
As seen from  the figure, 
the method is numerically exact, even at large interaction strengths, in sign-problem-free systems. 


\subsection{Spin-imbalanced system}
\begin{figure*}[ht]
	\begin{center}
		\includegraphics[width=\textwidth, angle=0]{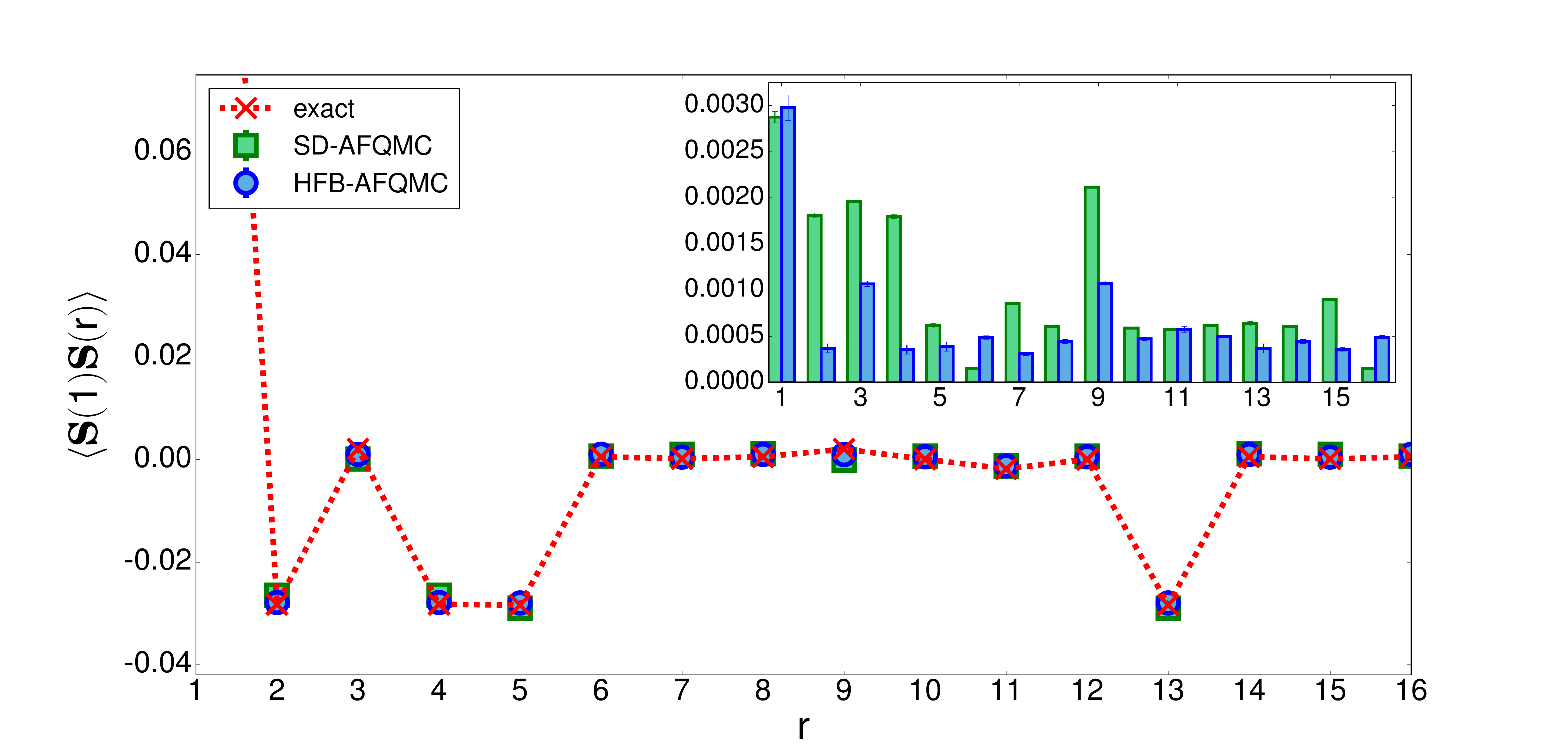}
		\caption{(Color online) Spin-spin correlation versus site number with comparison to ED. 
The system is a $4 \times 4$  lattice under periodic boundary conditions, with 
 $N_{\uparrow} = 7$ and $N_{\downarrow} = 5$ fermions and an interaction strength $U/t = -4$. 
		The results of two separate AFQMC calculations are plotted, the first with a Slater determinant trial wave function, and the second with a HFB (BCS) trial wave function. The inset plots the magnitude of the error (relative to the exact result) for each choice of trial wave function.	
		}
		\label{fig:spin-imbalanced:spin}
	\end{center}
\end{figure*}

We have demonstrated that a BCS trial wave function can yield numerically exact results in spin-balanced, sign-problem-free systems. In these systems the use of a BCS trial wave function 
can improve the statistical efficiency, and reduce the projection length in imaginary-time for 
reaching the ground state. 
We now consider the more computationally challenging case of non-zero spin polarization, which leads to 
the emergence of a sign-problem. 

When the sign-problem is present, the systematic accuracy, as well as the efficiency of the simulation, 
can be affected by the choice of trial wave function. 
Here we show that, for spin-imbalanced systems with a sign-problem, BCS trial wave functions obtained via the procedure outlined in Sec.~\ref{sec:discussion}, which makes use of a Hartree-Fock-Bogoliubov (HFB) transformation, offer a significant improvement over the standard choice of Slater determinants.

\begin{figure*}[ht]
	\begin{center}
		\includegraphics[width=\textwidth, angle=0]{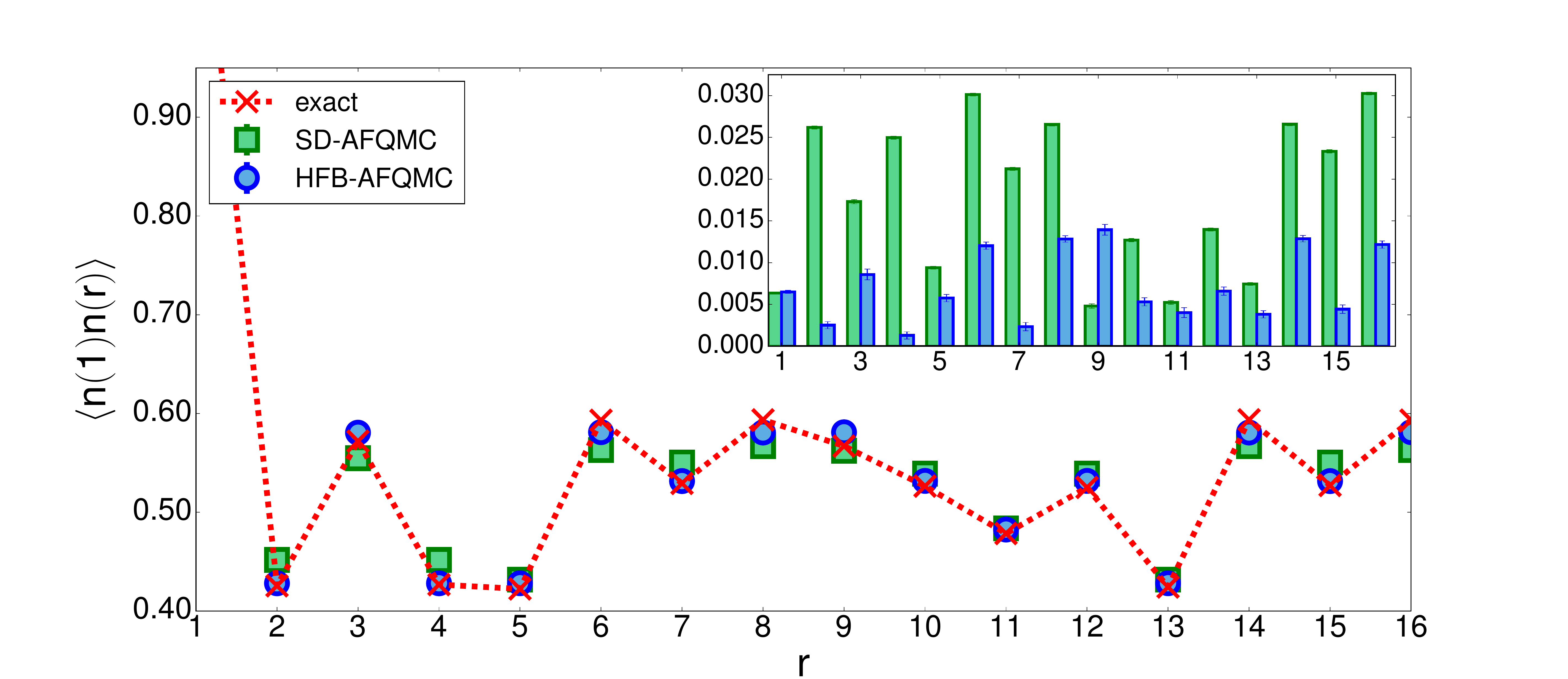}
		\caption{ 
			(Color online) Density-density correlation with comparison to ED. The inset shows the magnitude of the error as in Fig.~\ref{fig:spin-imbalanced:spin}.}
		\label{fig:spin-imbalanced:density}
	\end{center}
\end{figure*}
\begin{figure*}[ht]
\begin{center}
\includegraphics[width=\textwidth, angle=0]{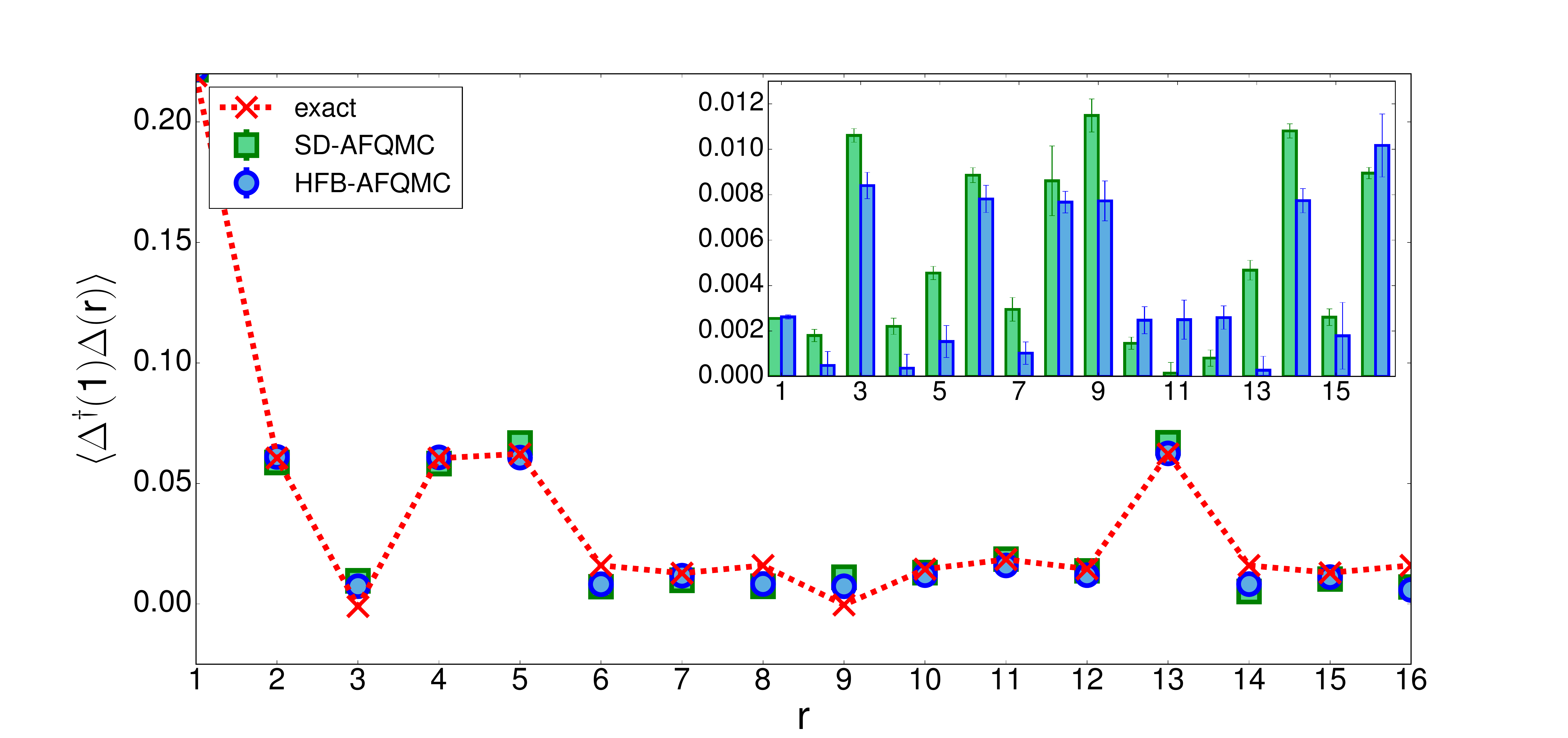}
\caption{ 
 (Color online) Pair-pair correlation with comparison to ED. The inset shows the magnitude of the error as in Fig.~\ref{fig:spin-imbalanced:spin}.}
\label{fig:spin-imbalanced:pairing}
\end{center}
\end{figure*}

\begin{figure*}[ht]
\begin{center}
\includegraphics[width=0.9\textwidth, angle=0]{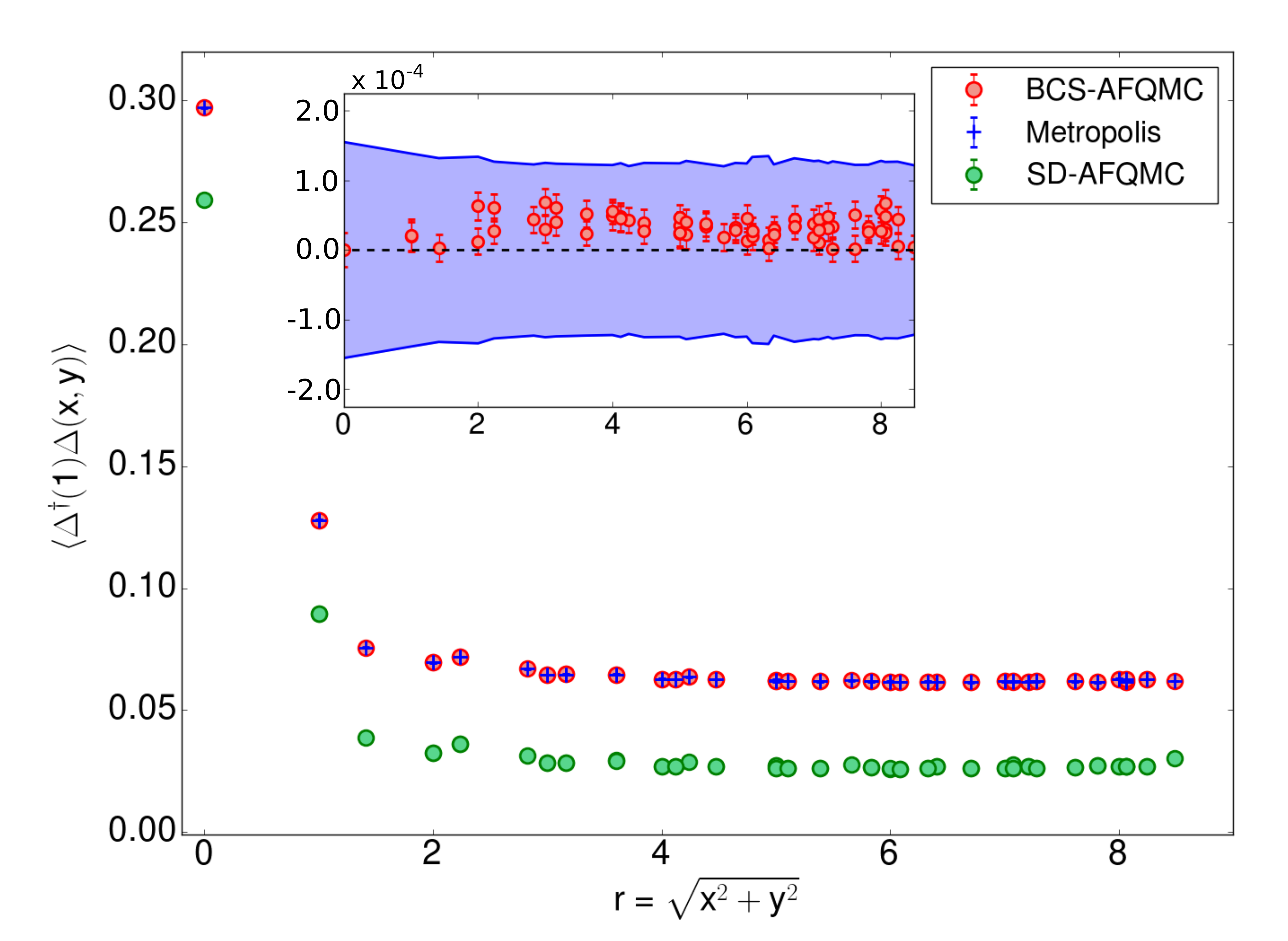}
\caption{ 
 (Color online) Pair-pair correlation of a system of $N=122$ particles, on a $12 \times 12$ lattice, at $U/t = -4$, with comparison to the path-integral Metropolis result. The inset plots the magnitude of the difference between the BCS-AFQMC result and the Metropolis result. The blue shaded represents the Metropolis error.}
\label{large:pairing}
\end{center}
\end{figure*}

\COMMENTED{
When the sign-problem is present, the accuracy and efficiency of the simulation is 
affected by the choice of trial wave function. We have already demonstrated that a BCS trial wave function can yield numerically exact results in spin-balanced, sign-problem-free systems. Here we show that for spin-imbalanced systems with a sign-problem, BCS trial wave functions obtained via the procedure outlined in Sec.~\ref{sec:discussion}, which makes use of a Hartree-Fock-Bogoliubov (HFB) transformation, offer a significant improvement over the standard choice of Slater determinants.
}

We provide an example of this improvement in Figs.~\ref{fig:spin-imbalanced:spin}, ~\ref{fig:spin-imbalanced:density}, and ~\ref{fig:spin-imbalanced:pairing}, which compare the accuracy of results for different observables calculated using AFQMC with a Slater determinant or BCS trial wave function. 
Results are shown for the case of $N_{\uparrow} =  7$, $N_{\downarrow} = 5$ on a two-dimensional lattice with $4 \times 4$ sites and periodic boundary conditions, at an interaction strength of $U/t = -4$, 
in which we can perform exact diagonalization for comparison.
For each observable, although AFQMC yields good results with qualitatively correct predictions 
of the correlation functions using either a single
determinant (SD) trial wave function or a projected BCS trial wave function derived from HFB, 
quantitative differences are evident.
The results obtained using the BCS trial wave function show better
agreement with the exact result. The insets in each figure, which plot the magnitude of the error relative to the exact result for either choice of trial wave function, show generally smaller errors for all three observables with the choice of BCS trial wave function.

\subsection{Large systems}
As a final example, we apply the new method  to a larger supercell to illustrate 
that it can scale straightforwardly and that it retains complete numerical stability for large system
size and long imaginary-time projection.
\COMMENTED{ 
Before moving to the discussion section, we find useful to show the results for a large system.} 
Figure~\ref{large:pairing} plots the on-site pairing correlation function for a $12 \times 12$ lattice, hosting $122$ particles with $U/t = -4$. Though this system is spin-balanced and therefore sign-problem-free, it can be used to test the constrained-path approximation, by allowing 
the force bias to naturally take effect and impose an artificial constraint. As we mentioned at the end of Sec.~\ref{sec:method}, when there is a sign problem 
(as opposed to a phase problem), the use of the force bias imposes a barrier at $\langle \psi_T|\phi^w\rangle=0$ such that, in the limit of $\delta \tau\rightarrow 0$, no walker would be able to cross.
In an actual application, once the absence of the sign problem is established (either with analytical 
arguments or empirical testing), we can simply add a small shift in the importance function, similar to  constraint 
release \cite{PhysRevB.88.125132}, to remove the barrier. But if we do not introduce this step,
the CP calculation would proceed, and a constraint bias can be present even if there is no 
sign problem. Here we take advantage of this feature as a way to both benchmark and illustrate 
the new method with BCS trial wave function.

\COMMENTED{
it is a well-known and extremely challenging test for the CP-AFQMC method with a Slater determinant trial wave function. 
The challenge arises primarily because the Slater determinant fails to capture the correct structure of the ground state wave function. Within the CP-AFQMC formalism the trial wave function is used to guide the random walk, i.e. the Monte Carlo sampling, and to prevent the exponential decay of the signal to noise ratio at large imaginary times in systems with a sign problem. As described in Sec.~\ref{sec:method} this is accomplished by eliminating walkers that cross the nodal surface defined by the trial wave function, as in (\ref{eqn:nodal_surface}). 
Consequently, the quality of the sampling of the ground state and the accuracy of the simulation is sensitive to the selection of trial wave function, even in a sign-problem-free simulation.
In the example we provide here, because the BCS trial wave function provides a better description of the pairing structure of the ground state wave function, compared to a standard Slater determinant, the sampling is significantly improved, which leads to a considerably more accurate representation of the ground state.
}

We use the path-integral formalism, as is commonly employed in sign-problem-free cases,
 which samples fixed length paths in auxiliary-field space using the Metropolis algorithm.
 The Metropolis sampling procedure can be accelerated using the idea of force bias,
 as introduced in Ref.~\cite{Hao-2DFG}, and the initial wave functions are chosen as 
 single determinants. 
Exact results are obtained with this approach to use as a benchmark here.
We then carry out two sets of  
CP calculations using the method discussed above, with either a single Slater determinant or
a projected BCS trial wave function.

As Fig.~\ref{large:pairing} reveals,  the CP-AFQMC calculation with Slater determinant trial wave function shows a bias, with the pairing correlation function strength considerably under-estimated.
The BCS trial wave function, on the other hand, enables the CP calculation to recover the exact result.
The tiny statistical error bars seen in the inset of Fig~\ref{large:pairing} are a reflection of the numerical stability and good efficiency of the new method as discussed earlier.

These results illustrate the potential of the new method for treating a variety of interesting systems 
where pairing correlation is important, for example spin-imbalanced atomic gas systems and 
optical lattices, which are experimentally accessible and which may 
 host exotic phases like FFLO states, as well as various model systems including 
 those relevant to unconventional superconductivity. Almost all of these systems would incur 
 a sign problem. The method we have introduced will allow us to use CP-AFQMC with 
 BCS trial wave functions to study these systems.

\section{Discussion}
\label{sec:discussion}
In the preceding sections we outlined a method for incorporating BCS trial wave functions into AFQMC simulations for computing observables and correlation functions, and demonstrated its implementation with a set of benchmark calculations. Here we address some additional technical aspects and comment on possible extensions and applications. 

\subsection{The choice of the BCS wave function}
\label{ssec:BCSwf-choice}

We begin by considering the construction of the BCS state to be used as trial wave function. 
The method we have introduced is independent of this discussion. However, we expect these 
considerations to be useful in applications for constructing the optimal trial wave functions. 

In the simple case of translationally invariant systems where singlet pairing with zero total momentum of the pair is expected, the formalism of standard BCS theory can be applied. The "textbook" definition of the fully-paired BCS wave function is:
\begin{equation}
|\Psi\rangle = \prod_{{\bf{k}}}\left( u_ {{\bf{k}}} + v_{{\bf{k}}} \hat{c}^{\dagger}_{{{\bf{k}}},\uparrow}
\hat{c}^{\dagger}_{{-{\bf{k}}},\downarrow}\right) \, |0\rangle
\label{standard_bcs}
\end{equation}
where $u_ {{\bf{k}}}$ and $v_{{\bf{k}}}$ are the coefficients of the celebrated Bogoliubov transformation.
When projected onto a sector of the Hilbert space 
with a fixed number of particles, the wave function \eqref{standard_bcs} can easily be recast in the form \eqref{sec1:projbcs}, with:
\begin{equation}
F^{}_{{\bf{r}},{\bf{r}}^{\prime}} = \frac{1}{L} \sum_{{\bf{k}}} \, e^{-i {{\bf{k}}} \cdot ({\bf{r}}-{\bf{r}}^{\prime})} \, \frac{v_{{\bf{k}}} }{u_ {{\bf{k}}}}
\end{equation}
where some care has to be taken if some of the coefficients $u_ {{\bf{k}}}$  vanish.
If there are unpaired fermions, their orbitals will be simple plane-waves.

More generally, the Hartree-Fock-Bogoliubov (HFB) methodology 
\cite{Ring_and_Schuck}
can be used in order to obtain a generalized Bogoliubov transformation of the form:
\begin{equation}
\left( \hat{\gamma}^{\dagger} \,\, \hat{\gamma}^{} \right)
= \, \left( \hat{c}^{\dagger} \,\, \hat{c}^{} \right) \,\, 
\left(
\begin{array}{cc}
U & V^{\star} \\
V & U^{\star}
\end{array}
\right)
\label{hartree-fock-bogoliubov}
\end{equation}
where the matrix of the transformation has $4L \times 4L$ elements. 
Here we show how to connect the transformation \eqref{hartree-fock-bogoliubov} 
to a wave function of the form \eqref{sec1:projbcs}. 
We build the unpaired orbitals, i.e. the matrix $D$, and the pairing matrix $F$ in such a way that the wave function \eqref{sec1:projbcs} is the vacuum
of the algebra of operators $\left\{ \hat{\gamma}^{\dagger} \,,\, \hat{\gamma}^{} \right\}$, that is:
\begin{equation}
\hat{\gamma}^{}_{i} \, |\psi_{BCS}\rangle = 0, \quad i=1,\dots,2L
\end{equation}


We first perform a singular value decomposition of the matrix $V$ in \eqref{hartree-fock-bogoliubov}:
\begin{equation}
V = \mathcal{A}^{\star} \, \mathcal{Z} \,  \mathcal{B}^{\dagger}\,,
\end{equation}
where $\mathcal{A}$ and $ \mathcal{B}$ are unitary complex matrices, while $\mathcal{Z} $ is real and diagonal, 
and introduce the $2L \times 2L$ complex matrix:
\begin{equation}
\label{eq:Wmat}
\mathcal{W} = \mathcal{A}^{\dagger} \, U^{}   \mathcal{B}\,.
\end{equation}
We next consider the density matrix and the pairing tensor, defined on the vacuum of the operators $\left\{ \hat{\gamma}^{\dagger} \,,\, \hat{\gamma}^{} \right\}$:
\begin{equation}
\begin{split}
& \rho_{ij} = \langle \hat{c}^{\dagger}_{j} \hat{c}^{}_{i} \rangle = \left(V^{\star} V^{T} \right)_{ij}   , \\ 
& \tau_{ij} =  \langle \hat{c}^{}_{j} \hat{c}^{}_{i} \rangle = \left(V^{\star} U^{T}\right)_{ij}\,.
\end{split}
\end{equation}
It is easy to see,
\begin{equation}
\rho   =  \mathcal{A} \, \mathcal{Z}^2   \,\mathcal{A}^{\dagger}\,,  
\end{equation}
meaning that the matrix $ \mathcal{A}$ is made of natural orbitals, which
we express as $\{ | \alpha \rangle, | \beta \rangle \dots \}$.
We have:
\begin{equation}
\mathcal{Z}^2_{\alpha,\beta} = \langle \hat{c}^{\dagger}_{|\beta\rangle} \hat{c}^{}_{|\alpha\rangle} \rangle =   \delta_{\alpha,\beta} \, \zeta_{\alpha}^2, \quad   \zeta_{\alpha}^2 \in [0,1]
\end{equation}
Similarly, 
\begin{equation}
\tau =  
\mathcal{A} \, \mathcal{Z} \, \mathcal{W}^{T} \, \mathcal{A}^{T} 
\end{equation}
and we have:
\begin{equation}
\left( \mathcal{Z} \, \mathcal{W}^{T}  \right)_{\alpha,\beta} = \zeta_{\alpha} \, \mathcal{W}_{\beta\alpha}
= \langle \hat{c}^{}_{|\beta\rangle} \hat{c}^{}_{|\alpha\rangle} \rangle 
\end{equation}
providing the interpretation of the elements of the matrix $\mathcal{W}$ introduced in Eq.~(\ref{eq:Wmat}).

\COMMENTED{
Now we go back to the transformation:
\begin{equation}
\left( \hat{\gamma}^{\dagger} \,\, \hat{\gamma}^{} \right)
= \, \left( \hat{c}^{\dagger} \,\, \hat{c}^{} \right) \,\, 
\left(
\begin{array}{cc}
U & V^{\star} \\
V & U^{\star}
\end{array}
\right)
\end{equation}
which is equivalent to the following one:
\begin{equation}
\left( \hat{\beta}^{\dagger} \,\, \hat{\beta}^{} \right)
= \, \left( \hat{a}^{\dagger} \,\, \hat{a}^{} \right) \,\, 
\left(
\begin{array}{cc}
\mathcal{W} & \mathcal{Z} \\
\mathcal{Z} &  \mathcal{W}^{\star}
\end{array}
\right)
\end{equation}
}

Now 
if we introduce the new operators
\begin{equation}
\left( \hat{\beta}^{\dagger} \,\, \hat{\beta}^{} \right) = \left( \hat{\gamma}^{\dagger} \,\, \hat{\gamma}^{} \right) 
\,\, 
\left(
\begin{array}{cc}
\mathcal{B}  & 0 \\
0 &  \mathcal{B}^{\star}
\end{array}
\right)
\end{equation}
and
\begin{equation}
\, \left( \hat{a}^{\dagger} \,\, \hat{a}^{} \right) \,\, 
= \, \left( \hat{c}^{\dagger} \,\, \hat{c}^{} \right) \,\, 
\left(
\begin{array}{cc}
\mathcal{A}  & 0 \\
0 &  \mathcal{A}^{\star}
\end{array}
\right)\,,
\end{equation}
we see that the transformation in Eq.~(\ref{hartree-fock-bogoliubov}) is equivalent to the following:
\begin{equation}
\left( \hat{\beta}^{\dagger} \,\, \hat{\beta}^{} \right)
= \, \left( \hat{a}^{\dagger} \,\, \hat{a}^{} \right) \,\, 
\left(
\begin{array}{cc}
\mathcal{W} & \mathcal{Z} \\
\mathcal{Z} &  \mathcal{W}^{\star}
\end{array}
\right)\,.
\end{equation}
%
The unitarity of the matrix on the right-hand side above implies that
\COMMENTED{
\begin{equation}
\left(
\begin{array}{cc}
\mathcal{W} & \mathcal{Z} \\
\mathcal{Z} &  \mathcal{W}^{\star}
\end{array}
\right)
\end{equation}
}
\COMMENTED{
	must be unitary:
	\begin{equation}
	\label{uniconditions}
	\begin{split}
	& \left(
	\begin{array}{cc}
	\mathcal{W} & \mathcal{Z} \\
	\mathcal{Z} &  \mathcal{W}^{\star}
	\end{array}
	\right) \, \left( \begin{array}{cc}
	\mathcal{W}^{\dagger} & \mathcal{Z} \\
	\mathcal{Z} &  \mathcal{W}^{T}
	\end{array}
	\right) = \mathbb{I} \\
	& \left( \begin{array}{cc}
	\mathcal{W}^{\dagger} & \mathcal{Z} \\
	\mathcal{Z} &  \mathcal{W}^{T}
	\end{array}
	\right) \,
	\left(
	\begin{array}{cc}
	\mathcal{W} & \mathcal{Z} \\
	\mathcal{Z} &  \mathcal{W}^{\star}
	\end{array}
	\right)
	= \mathbb{I}
	\end{split}
	\end{equation}
}
\begin{equation}
\left(   \mathcal{W}^{\dagger}   \mathcal{W} \right)_{\alpha,\beta} = \left(   \mathcal{W}   \mathcal{W}^{\dagger} \right)_{\alpha,\beta} = \delta_{\alpha,\beta}
\left( 1 - \zeta_{\alpha}^2 \right)\,,
\end{equation}
considering that $\mathcal{Z}$ is real and diagonal.
In other words, the rows and columns of the matrix $\mathcal{W} $ are orthogonal vectors,
their squared norm is $\left( 1 - \zeta_{\alpha}^2 \right)$, and they are identically zero whenever $\zeta_{\alpha}^2 = 1$.

We write explicitly:
\begin{equation}
\begin{split}
&
\hat{\beta}^{\dagger}_{l} = \sum_{ k} \, \hat{a}^{\dagger}_{k} \mathcal{W}_{kl}
+  \zeta_{l} \hat{a}^{}_{l} \\
&  \hat{\beta}^{}_{l} = \zeta_{l} \hat{a}^{\dagger}_{l}  + \sum_{ k} \, \hat{a}^{}_{k} \mathcal{W}^{\star}_{kl}\,.
\end{split}
\end{equation}
We have two possible situations that lead us to classify the states as occupied ("o") or paired ("p"):
\begin{equation}
\begin{split}
& \zeta^2_{o} = 1   \Rightarrow
\begin{cases}
\hat{\beta}^{\dagger}_{o} =  \zeta_{o} \hat{a}^{}_{o} \\
\hat{\beta}^{}_{o} = \zeta_{o} \hat{a}^{\dagger}_{o}  
\end{cases} \\
&  \zeta^2_{p} \in [0,1)   \Rightarrow
\begin{cases}
\hat{\beta}^{\dagger}_{p} = \sum_{ k} \, \hat{a}^{\dagger}_{k} \mathcal{W}_{kp}
+  \zeta_{p} \hat{a}^{}_{p}  \\
\hat{\beta}^{}_{p} = \zeta_{p} \hat{a}^{\dagger}_{p}  + \sum_{ k} \, \hat{a}^{}_{k} \mathcal{W}^{\star}_{kp}
\end{cases}
\end{split}
\end{equation}


We define $\tilde{\mathcal{W}}$ to be the matrix obtained from $\mathcal{W}$
keeping only the rows and columns corresponding to the $p$ orbitals, for which $ \zeta^2_{p} \in [0,1)$. We similarly define $\tilde{\mathcal{Z}}$. 
With the preceding definitions we construct the antisymmetric matrix:
\begin{equation}
\mathcal{O} = \left( \tilde{\mathcal{Z}} \tilde{\mathcal{W}}^{-1}   \right)^{\star}
\end{equation}
and the wave function:
\begin{equation}
\left| \psi_{BCS} \right\rangle = \mathcal{N} \, \left( \prod_{o} \hat{a}^{\dagger}_{o} \right)\,\, e^{\frac{1}{2}  \sum_{p,p^{\prime}} \mathcal{O}_{p,p^{\prime}} \, \hat{a}^{\dagger}_{p} \hat{a}^{\dagger}_{p^{\prime}}   } \, | 0 \rangle
\end{equation}
where $o$ runs only over the occupied states, while $p$ and $p^{\prime}$ run over the paired states. The pre-factor $\mathcal{N}$ is included for normalization. This wave function is the BCS wave function we set out to construct.
\COMMENTED{
If we introduce the wave function:
\begin{equation}
\left| \psi_{BCS} \right\rangle = \mathcal{N} \, \left( \prod_{o} \hat{a}^{\dagger}_{o} \right)\,\, e^{\frac{1}{2}  \sum_{p,p^{\prime}} \mathcal{O}_{p,p^{\prime}} \, \hat{a}^{\dagger}_{p} \hat{a}^{\dagger}_{p^{\prime}}   } \, | 0 \rangle
\end{equation}
where $o$ runs only over the occupied states, while $p$ and $p^{\prime}$ run over the paired states, and we define the antisymmetric matrix:
\begin{equation}
\mathcal{O} = \left( \tilde{\mathcal{Z}} \tilde{\mathcal{W}}^{-1}   \right)^{\star}
\end{equation}
then $\left| \psi_{BCS} \right\rangle$ is the BCS wave function we are looking for.}

It is a simple exercise to verify that 
\COMMENTED{
	We will show, in fact, that:
	\begin{equation}
	\hat{\beta}_l \, \left| \Psi_{BCS} \right\rangle = 0, \quad \forall l
	\end{equation}
	which implies that:}
\begin{equation}
\hat{\gamma}_l \, \left| \psi_{BCS} \right\rangle = 0, \quad  l=1,\dots,2M
\end{equation}
\COMMENTED{
	since the transformation between $\beta$ and $\gamma$ does not mix creators and destructors.
	$ \left| \Psi_{BCS} \right\rangle $ is thus the vacuum of the Bogoliubov operators.
}
In order to go back to the original basis of lattice sites, we simply need to use the transformation
\begin{equation}
\hat{a}^{\dagger}_{i}  = \sum_{{\bf{r}},\sigma} \hat{c}^{\dagger}_{{\bf{r}},\sigma} \mathcal{A}_{({\bf{r}},\sigma),i}, \quad i=p,o
\end{equation}
\COMMENTED{
	which implies:
	\begin{equation}
	\begin{split}
	& \sum_{p,p^{\prime}} \mathcal{O}_{p,p^{\prime}} \, \hat{a}^{\dagger}_{p} \hat{a}^{\dagger}_{p^{\prime}} 
	=  \sum_{{\bf{r}},\sigma \, , \, {\bf{r}}^{\prime},\sigma^{\prime} }
	\sum_{p,p^{\prime}}  \mathcal{A}_{({\bf{r}},\sigma),p} \, \mathcal{O}_{p,p^{\prime}} \, 
	\mathcal{A}_{({\bf{r}}^{\prime},\sigma^{\prime}),p^{\prime}}
	\end{split}
	\end{equation}
}
In the case of singlet pairing, $\left| \psi_{BCS} \right\rangle $ can be immediately recast in the form
\eqref{sec1:projbcs}, with:
\begin{equation}
\label{sec4:matD}
D_{{\bf{r}},i=o} = \mathcal{A}_{({\bf{r}},\uparrow),o} 
\end{equation}
and:
\begin{equation}
\label{sec4:matF}
F_{{\bf{r}},{\bf{r}}^{\prime}} =  \sum_{p,p^{\prime}}  \mathcal{A}_{({\bf{r}},\uparrow),p} \, \mathcal{O}_{p,p^{\prime}} \, 
\mathcal{A}_{({\bf{r}}^{\prime},\downarrow),p^{\prime}}
\end{equation}
The relations \eqref{sec4:matD} and \eqref{sec4:matF} outline as 
algorithm that provides an interface between the HFB calculation and the QMC simulation. 
This interface allows us to use a wave function obtained from an HFB transformation as a trial wave function for a CP-AFQMC calculation. We also note that the same HFB transformation can be used to construct an initial wave function that has good overlap with the trial wave function. 

\COMMENTED{
\subsection{Computational cost}

The framework we have developed is a computationally efficient approach to the calculation of ground state observables. To better illustrate this efficiency, we perform a comparison of computational cost between AFQMC simulations using SD and BCS trial wave functions.  
}
\subsection{Possible extensions} 
\label{ssec:extensions}

While the methodology we have described focuses on the use of BCS states in the constrained-path formalism, the same technique can be applied to wave functions of different forms. This includes multi-determinant trial wave functions, especially when they are composed of non-orthogonal Slater determinants. Multi-determinant wave functions are of considerable importance in quantum chemistry calculations \cite{Mario_AFQMC_QC, Morales_QMC_multi}, and have been shown to provide high-accuracy results. 

This technique can be implemented in the path-integral AFQMC framework, in which fixed-length paths in auxiliary-field space are sampled using the Metropolis algorithm. Path-integral AFQMC has a long record of success in the treatment of various model Hamiltonians, including those with exotic pairing behaviors, such as the attractive Fermi gas \cite{AFQMC_BCS,Bulgac_QMC_BCS-BEC,Hao-2DFG,2DFG_Drut,soc-qmc,AFQMC_Rashba_OPLATT}. 
The use of BCS wave functions would help extend the reach of the method and enable the simulation of even larger systems, with important pairing properties, in both two and three dimensions. 
One can use a BCS trial wave function at one end of the path, while using either a single-/multi-determinant trial wave function or a BCS trial wave function on the other. (If
we use BCS trial wave function on both ends of the path, we could alternatively view the formalism 
as propagating in HFB space, which has been discussed in Ref.~\cite{Hao-HFB-PhysRevB.95.045144}.)

Similarly, the same technique we have discussed can be used in a mean-field context, 
which can be considered a specialized case of the AFQMC, with only a single path instead 
of the path integral. For example, in a mean-field calculation formulated as an imaginary-time 
projection \cite{Wirawan-PRE-2004}, exactly the same formalism can be adopted. The technique may also be useful 
for time-dependent mean-field calculation for dynamical properties.

\section{Conclusions}

We have presented 
a method for computations of observables and correlation functions 
using a BCS state as a trial wave function in many-body computations involving path integrals in Slater determinant space.
We illustrated the method with a set of benchmark calculations comparing with existing technology
and also in situations where a sign problem is present, by comparison with exact diagonalization.
We demonstrated that the method removes any numerical instabilities in propagating BCS 
states. The method controls the sign problem and has computational cost scaling 
as a low power with system size.

 For attractive interactions, the methodology  provides a clear improvement over the cutting-edge CP-AFQMC technique relying on Slater determinant trial wave functions. This development will enable high-accuracy computations
 of exotic superfluid phases, like FFLO states. 
 In the realm of repulsive models and molecular/solid systems, this approach will allow the direct use of a pairing trial wave function in 
AFQMC calculations, which can be advantageous in, for example, the study of 
models for interacting electrons where pairing is expected. 
%
 
 The work we have presented here can serve as a general framework for incorporating BCS and other beyond-Slater determinant wave functions into quantum Monte Carlo calculations working in second quantization. 
 The formalism introduced, namely to replace back-propagation in the computation 
of observables and correlation functions by forward propagating the corresponding Green function 
matrix, can be useful in  other contexts. 
 We hope these developments will enable many applications in a variety of problems, and also
 stimulate further methodological improvements 
 in the study of strongly correlated models.

We thank J.~Carlson, A.~Gezerlis, Lianyi He, and Hao Shi for helpful discussions.
This work was supported by 
NSF (Grant No. DMR-1409510) and by the  
Simons Foundation. Computing was carried out at the Extreme Science and Engineering Discovery Environment (XSEDE),
which is supported by National Science Foundation grant number ACI-1053575, and the computational
facilities at William and Mary. 
The Flatiron Institute is a division of the Simons Foundation.


%


\end{document}